\documentclass[]{spie}  

\usepackage{amsmath,amsfonts,amssymb}
\usepackage{graphicx}
\usepackage[colorlinks=true, allcolors=blue]{hyperref}

\title{Modal noise in few-mode fibers}
\authorinfo{Further author information, send correspondence to N.Blind (nicolas.blind@unige.ch)}

\author[a]{N. Blind}
\affil[a]{Observatoire astronomique de l'Universit\'e de Gen\`eve, 51 Chemin de Pegasi, CH-1290 Sauverny, Switzerland}

\begin{document} 
\maketitle

\begin{abstract}
NIRPS (Near Infra-Red Planet Searcher) is an AO-assisted and fiber-fed spectrograph for high precision radial velocity measurements in the YJH-bands. NIRPS also has the specificity to be an SCAO assisted instrument, enabling the use of few-mode fibers for the first time. This choice offers an excellent trade-off by allowing to design a compact cryogenic spectrograph, while maintaining a high coupling efficiency under bad seeing conditions and for faint stars. The main drawback resides in a much more important modal-noise, a problem that has to be tackled for allowing 1m/s precision radial velocity measurements.

We present in this paper the result of a semi-empirical work that allowed to validate the scrambling device and strategies to mitigate modal noise. It is based at first on a complete set of lab measurements of the final fibers. Second, such measurements are injected in the spectrograph design to study in particular the impact of grating and optics illumination on derived RVs.

\end{abstract}


\section{INTRODUCTION}
\label{sec:intro}  

Since its 1st light in 2002, HARPS has been setting the standard in the exo-planet detection by precision radial velocity (pRV) measurements. Based on this experience, our consortium is developing a high accuracy near-infrared RV spectrograph covering YJH bands to detect and characterize low-mass planets in the habitable zone of M dwarfs. It will allow RV measurements at the 1-m/s level and will look for habitable planets around M- type stars by following up the candidates found by the upcoming space missions TESS, CHEOPS and later PLATO. Although several similar instruments are already operated (CARMENES, GIANO, SPIROU), NIRPS is the only one of its kind in the Southern hemisphere. NIRPS and HARPS, working simultaneously on the ESO 3.6m will become a single powerful high-resolution spectrograph covering from 0.4 to 1.8 micron. NIRPS will complement HARPS in validating earth-like planets found around G and K-type stars whose signal is at the same order of magnitude than the stellar noise. A complete description of the system can be found in \cite{wildi_2017a}.

Being an AO-assisted fiber-fed NIR spectrograph, NIRPS fibers works in a so-called few-mode regime, that is in-between the usual multi-mode regime (where 1000s of modes propagate) and the single-mode. Although few-mode fibers are prone to higher modal noise, we present in this paper our work to characterize and manage it. We show that it shall not preclude the pRV performance to be achieved. This choice nevertheless allowed for designing a much more compact spectrograph and grating (with a linear dimension gain of $\sim$2.5), a serious advantage in the NIR, where the instrument must be cooled.

\subsection{Few-mode fibers coupling}
The number of modes that can transport a circular waveguide is roughly equal to $V^2 /4$, where V is the fiber parameter: %
\begin{equation}
V = 2 \pi\: \mathrm{NA} \: a / \lambda
\end{equation}
where $NA$ is the fiber numerical aperture, $a$ the core radius and $\lambda$ the wavelength. The NIRPS High Accuracy fiber (Sect.~\ref{sec:haf}) has a core size diameter of 29$\mu m$ (0.4" on sky) and only transports 10 to 35 modes, hence working in the so-called few-mode regime. This must be compared to the standard seeing-limited, visible fibers containing 1000s of modes. In this regime, geometrical optics does not apply so easily anymore, regarding coupling $\rho$ or scrambling properties.

Thanks to partial AO correction, a fiber of size 0.4" on-sky can reach coupling $\rho$ of up to 70\% in median seeing condition on a bright target (and $\rho \ge 50\%$ for I$\le 12$) \cite{blind_2017a}. Such fibers can couple light within 0.5" in this regime, because modes are not totally constrained into the fiber core but are guided through a small section of the fiber cladding. While AO-assisted spectrograph generally push towards the use of modal-noise-free single-mode fibers, such fibers are very sensitive to wave-front errors and AO correction. Few-mode fibers allow to relax constraints on optical design as well as on the AO, allowing fainter limiting magnitude. As a rule of thumb, it appears that 5\% coupling loss is achieved for low-order aberrations amounting to 100\:nm RMS.

Modal noise is the main drawback of using few-mode fibers. Because of the limited number of modes (a.k.a. speckles in the multi-mode case), each has a stronger wait in pRV stability when they evolve, due to change in injection conditions, external stress, temperature drifts, etc. Such fibers also have much higher inter-modal dispersion, making inter-modal coupling (i.e. exchange of energy between modes) more difficult: the usual scrambling strategies, like fiber agitation, is a lot less effective. This also makes the far-field a lot more dependent on the near-field properties, which led us to question the efficiency of the double scrambler (Sect.~\ref{sec:double_scrambler}).

In this paper, we focus on the tests and results obtained for the final fibers of NIRPS. We do not present the many preliminary stability and FRD tests we performed on a larger octagonal fiber ($50\mu m$ core), or a smaller round fiber ($25 \mu m$ core). Although we performed analysis at constant V-number (i.e. shorter wavelength for larger fibers), to keep modal properties as close as possible from the final ones, the results are not 100\% transposable to the final fiber.

\begin{figure}[t]
    \centering
    \includegraphics[width=0.95\textwidth]{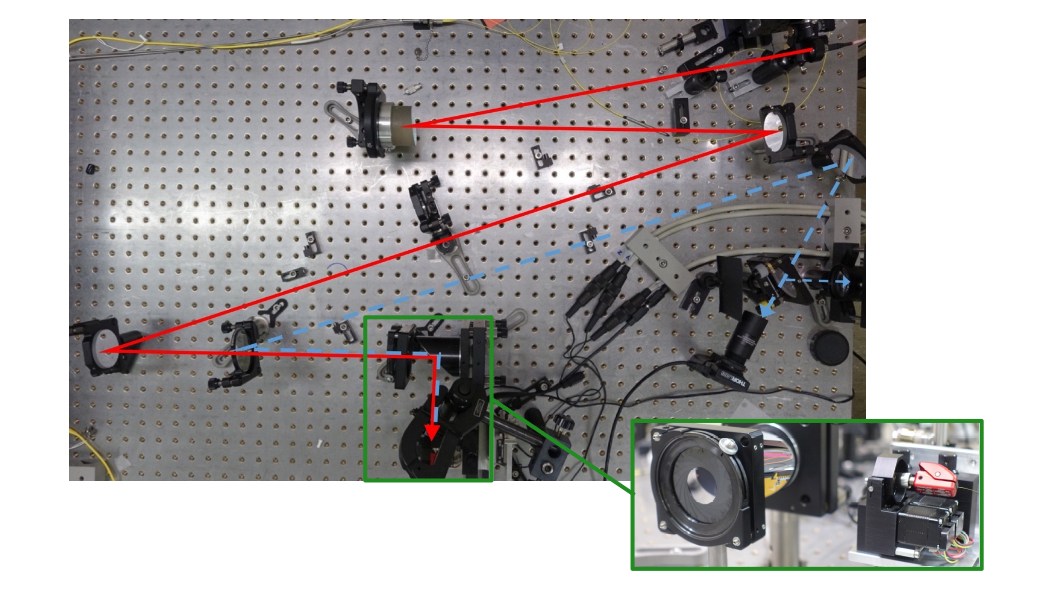}
    \caption{Pictures of NIRPS fiber bench. Light for testing fibers is shown in red, from top right to bottom. The blue optical path is the back-reflection alignment path, allowing bench focus and pupil imaging, and fiber near- and far-field alignment. In green, we zoom on aperture stop, exit OAP, and the 5 axis alignment stages.}
    \label{fig:bench}
\end{figure}

\subsection{NIRPS fibers}
We give here a short description of the fiber-link fibers. More details can be found in \cite{bouchy_2022a} (these proceedings).

\subsubsection{High accuracy fiber (HAF) - OCT29}
\label{sec:haf}
The standard high resolution fiber of NIRPS is a custom octagonal fiber from Ceramoptec, with core and cladding size of 29$\mu m$ and 116 $\mu m$ respectively, and NA=0.22. We measured its core size to 29 $\pm$ 0.3 $\mu m$. The core size corresponds to 0.4" on sky, and it is fed with an f/4.2 beam. It is used for the main science fiber (HAF), as well as for the 2 simultaneous reference / sky fiber. It allows for spectral resolution of R$\sim$100.000.

\subsubsection{High efficiency fiber (HEF) - OCT66 \& REC33x132}
\label{sec:hef}
The HEF fiber is a seeing-sized fiber of 0.9", planned to be used on fainter targets, when the AO cannot perform optimally. At the front-end injection level, it is a Ceramoptec octagonal fiber with core / cladding size 66$\mu$ / 264$\mu m$ respectively (drawn from the same preform as the OCT29), and fed with f/4.2 beam. In the double scrambler, a Bowen-Walraven slicer cuts the far-field in 2 halves, reimaged on a rectangular fiber of size 33x132 $\mu m$, also from Ceramoptec. Hence the rectangular fiber is illuminated on its whole surface in a pretty constant way. This slicing operation (which was meaningful thanks to the large size of the focal plane detector) allowed to preserve a rather high spectral resolution R$\sim$80.000.

\section{Method}

We adopted a semi-empirical approach to study the performance of the few-mode fibers and the different scrambling device. The fibers and scrambling device are first characterized on a bench, in the Near-Field (NF) and Far-Field (FF). 

Because of the few-mode nature of this fiber, in-fiber far-field scrambling is rather poor and double scrambler proves to be of limited interest (see Sect.~\ref{sec:double_scrambler}). Therefore, the far-field illumination of the grating and optics can significantly vary with injection conditions (AO guide star magnitude, seeing, etc). This effect cannot be studied simply by near-field measurement on a bench. The previous measurements were therefore also injected into the Zemax optical model of the spectrograph to study the impact of optical aberrations when pupil/grating illumination varies.

\begin{figure}[t]
    \centering
    \includegraphics[width=0.8\textwidth]{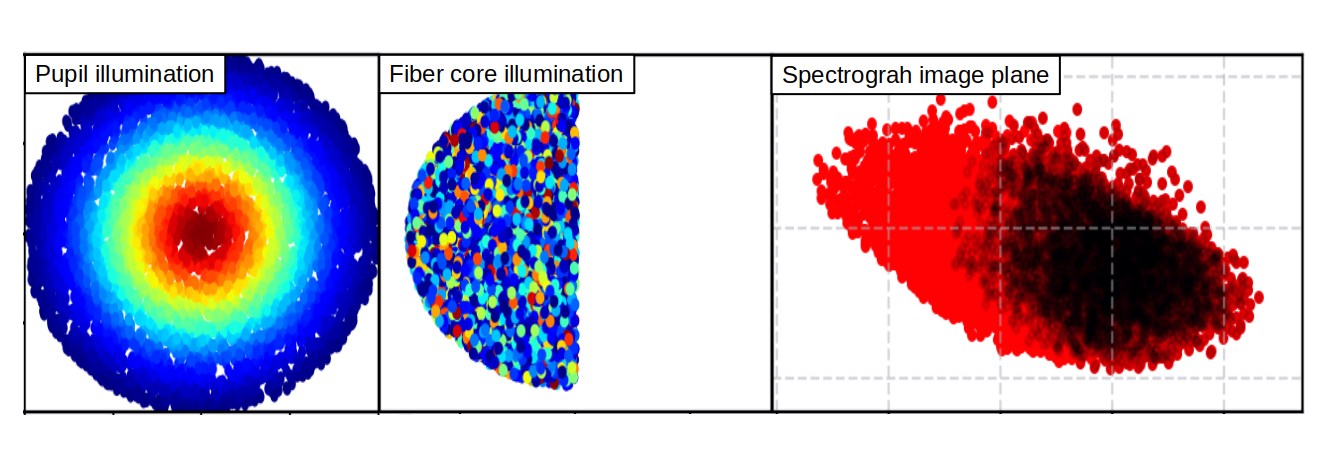}
    \caption{Example of spectrograph simulation with only 20.000 rays here. Left and middle: rays, colored by intensity in the pupil (weight). Half the fiber core is hidden in this example. Right: Resulting image at the focal plane at a particular wavelength (black). Red is the image formed using all rays (covering the fiber core and the complete grating).}
    \label{fig:spectro_rays}
\end{figure}

\subsection{Fiber test bench}
We set-up an achromatic bench (Fig.~\ref{fig:bench}), made of 2 OAPs and two folding mirrors (initially planed to allow for integrating the NIRPS AO system). The magnification of the bench is x0.2. While the input is fixed, the output is motorized, with 3-axis linear piezostage from Attocube, and a 2 rotational axis Gimbal mount from Zaber. The exit aperture can be varied from f/2.5 to less than f/22 for Focal Ratio Degradation (FRD) measurements. Apart from such tests the bench was used at f/4.2, the NIRPS fiber aperture. A pellicle beamsplitter allows to visualize the bench exit focal plane and pupil via back reflection, for fiber alignment via retroinjection. 

The fibers under test are mounted on the 5-axis motorized stage. For each, the procedure consists in recording NF and FF data for multiple positions across the core, typically leading to more than 200 measurements. We generally use an halogen lamp as primary light source, and have at disposal different narrow band filters ($\Delta\lambda\sim 10nm$), centered at 908, 1200, 1310, 1400, 1550 and 1650 nm. 

The bench has proven a stability better than 1m/s over time scale of 10-20 minutes, necessary to perform a complete XY scan of the fiber tip.

\subsection{Spectrograph simulations}
\label{sec:spectro_sim}
To study the behavior of the spectrograph in varying conditions, we first used its Zemax model to randomly generate 1.000.000 rays that uniformly cover the input focal (X/Y) and pupil planes (angle $\alpha$/$\beta$), and record their arrival position on the detector. We repeat the operation for 5 wavelengths along each order of the spectrograph, and for each of the 69 orders (Fig.~\ref{fig:spectro_rays}). The same set of rays was used in all cases to not add noise from the random sampling.

We then load those rays in a python script, with which we are able to reproduce the near- and far-field intensity measured on the bench by playing on the weight of each ray. One limitation is of course the broadband nature of our measurements ($\Delta\lambda=10nm$). To cover the full spectral range of the spectrograph, we  adopted two extrapolation methods:
\begin{enumerate}
    \item We take each measurement sets and make a linear interpolation of the profile with wavelength;
    \item We average the profiles of the different sets and inject it as an achromatic illumination.
\end{enumerate}
Both methods showed rather similar behavior.

\begin{figure}[t]
    \centering
    \includegraphics[width=1\textwidth]{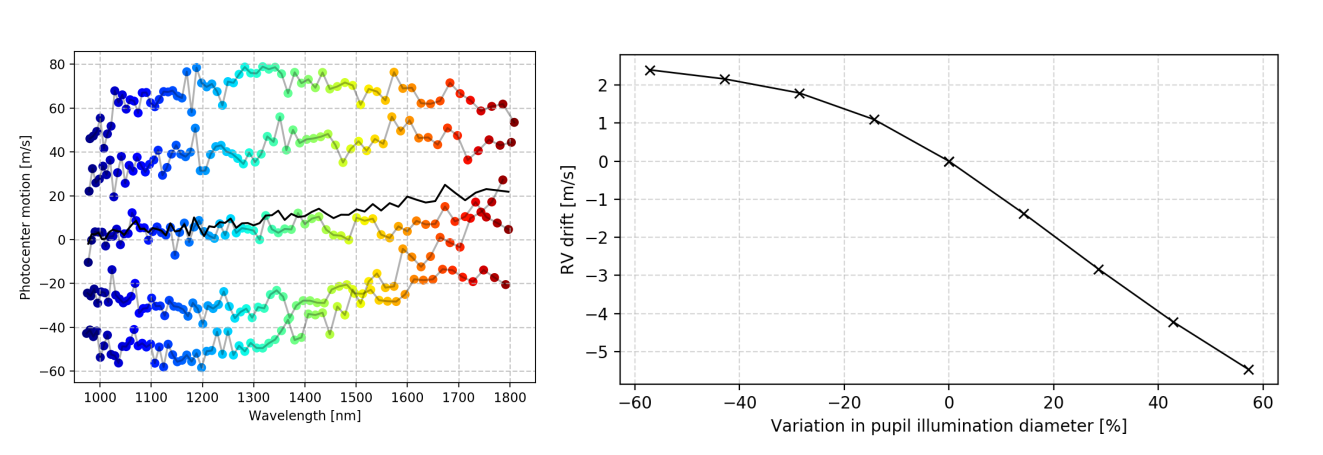}
    \caption{Impact of a change in the grating illumination on the RV value. Left: illustrative example of RV offset along all orders, with 5 points per order (dots, colored proportionnal to wavelength), and RV averaged over each order (black line). A line between the color dots would have a sawtooth shape. Right: summary of average RV offset (average of black curve in left plot) when varying the grating FWHM illumination for a fiber far-field with gaussian shape.}
    \label{fig:spectro_behavior}
\end{figure}

We also made more fundamental studies. As an example, we changed the grating (pupil) illumination, e.g. by varying the fiber far-field illumination size in Fig.~\ref{fig:spectro_behavior}, considering a theoretical gaussian illumination of the grating, and a fixed near-field illumination. This test demonstrates that far-field variations generates significant RV drift along every single order, but the symmetrical design allows to balance from one side to the other of each orders, so that average drift is nearly 0.

\section{Scrambling performance of the HAF / OCT29 fiber}

The HAF fiber was identified as the most problematic fiber, especially regarding the AO-corrected injection, and was therefore thoroughly studied.

\subsection{Natural behavior}

\begin{figure}
    \centering
    \includegraphics[width=1.0\textwidth, trim={0 0.5cm 0 0},clip]{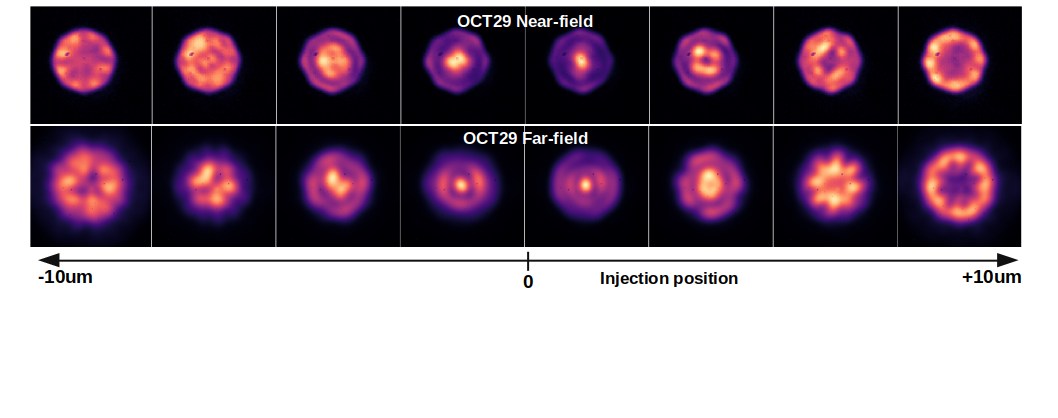}
    \caption{A few images of the OCT29 fiber NF and FF illumination. From left to right the fiber entrance is scanned with a point source, with 1310nm filter. }
    \label{fig:oct29_field}
\end{figure}

Fig.~\ref{fig:oct29_field} shows near- and far-field of a single OCT29 fiber at $\lambda$=1600nm, when scanning the entrance with a point source. Conversely to multi-mode fibers where the scrambling properties allow clear distinction between near- and far-fields, this is much more difficult here: there is a very strong correlation between near- and far-field illuminations.  This is also indicative that in the case of NIRPS, stabilizing the spectrograph pupil illumination is equivalent to stabilizing the fiber near-field. We discuss the interest of the double scrambler in Sec.~\ref{sec:double_scrambler}.

When scanning the fiber tip of the OCT29 fiber alone, we observe a photocenter motion of the order of 10m/s RMS (Fig.~\ref{fig:oct29_field} (top) and Fig.~\ref{fig:oct29_1600}) with narrow band filters. FRD of the fibers is pretty good, with 90\% at f/4.2, although the definition of FRD might be questionable given the extension of the modes (Fig.~\ref{fig:oct29_field}, bottom). The RV 'noise' for varying injection positions is presented in Fig.~\ref{fig:scrambling_summary}, together with performance after considering some of our scrambling device. The stability is computed for all injection position presenting a coupling higher than 50\% of the maximum. 

\subsection{Scrambling with adaptive optics}
\label{sec:AO}

Given the presence of an AO system in NIRPS, it is only natural to consider using the DM for scrambling. We demonstrated in \cite{blind_2017a} that mode-selective coupling is possible to some degree, by playing on the DM phase to match the phase of some specific modes. But the coupling losses are generally unacceptable. It appeared that the most efficient scrambling strategy with the AO of NIRPS is, by far, tip-tilt scan of the fiber tip: it preserves coupling into fiber and is able to couple to any given mode of the fiber, conversely to other classical optical aberrations. For instance, aberrations with azimutal symmetry like focus or sphericity are the worst aberrations: due to their symmetric nature, those terms reduce coupling in all modes, but are not able to excite new ones.

In practice, the AO-corrected PSF is describing a rose curve generated by updating the reference slopes at each AO loop iteration. The radial position is scanned with sine or triangular modulation at $\sim$20\:Hz, while the azimutal modulation is performed at a constant angular speed. It is possible to limit the capability of the fiber to hit the most central part of the fiber and to describe a ring pattern instead of a full disk.

\begin{figure}
    \centering
    \includegraphics[width=.95\textwidth]{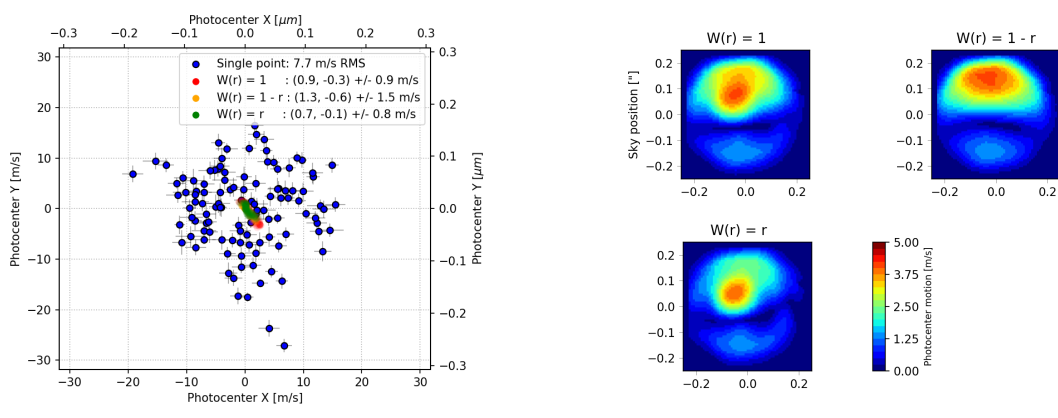}
    \caption{NF photocenter stability for a  single OCT29 at 1650nm. Left: scatter plot of 'natural' photocenter positions (blue, with error bars), as well as 'AO scrambling' averaged for different weighting strategies (red, orange, green). Right (3 images): representation of the photocenter instability for the 3 modulations, for various guiding offset ('Sky position' represents the center of the modulation).}
    \label{fig:oct29_1600}
\end{figure}

The measurements on the fiber bench are performed with a diffraction-limited source. The collect of scanned data allows a-posterior averaging of NF images to simulate different scrambling strategies (Fig.~\ref{fig:oct29_1600}). We compared 3 basic weighting laws across the fiber tip: 'uniform' (triangular modulation), 'r' (more weight in the outside of the fiber core, close from sine modulation) or '1-r' (more weight in the center of the fiber core). From one measurement to the other or different filter / wavelength band, only the '1-r' shows a slightly lower performance. We notice RV residuals are naturally much lower than without this scanning, allowing for fiber centring error of the order of $\pm$ 0.1" to keep $\le$ 0.5m/s NF stability.

\begin{figure}
    \centering
    \includegraphics[width=0.95\textwidth]{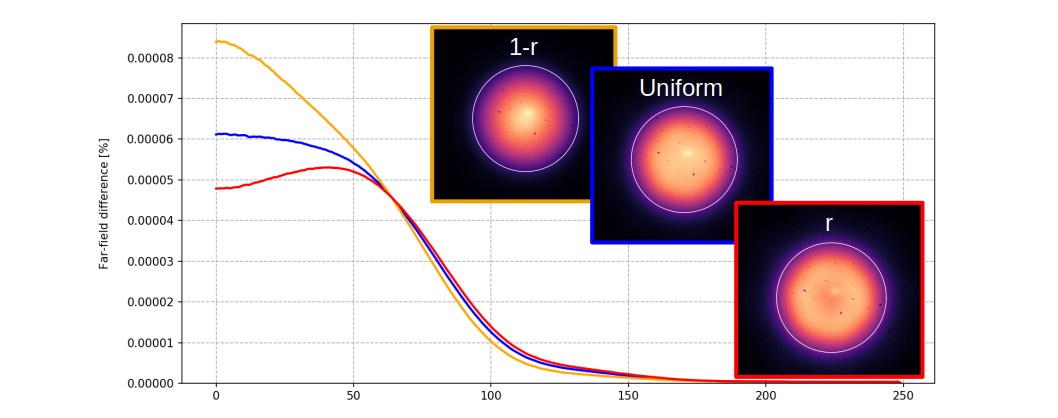}
    \caption{Far-field illumination and profile cut resulting from AO scrambling with different laws at 1650nm. f/4.2 corresponds to pixel $\sim$100}
    \label{fig:oct29_FF_scrambling}
\end{figure}

Measurements performed in the far-field show how the scrambling strategy can influence the grating illumination (Fig.~\ref{fig:oct29_FF_scrambling}), primarily the center of the far-field illumination, and almost not the external. The uniform law tends to deliver flatter far-field over 50\% of the fiber NA cone (but this is does not constitute a criterion). Very similar profiles are observed with the other filters, down to 1000 nm. This behavior is nevertheless observed only in the case of a single fiber. Far-field illumination with the stretcher is much less dependent on the scrambling choice thanks to internal scrambling (see Sect.~\ref{sec:stretcher}).

\subsection{Stretcher}
\label{sec:stretcher}

\begin{figure}
    \centering
    \includegraphics[width=0.8\textwidth]{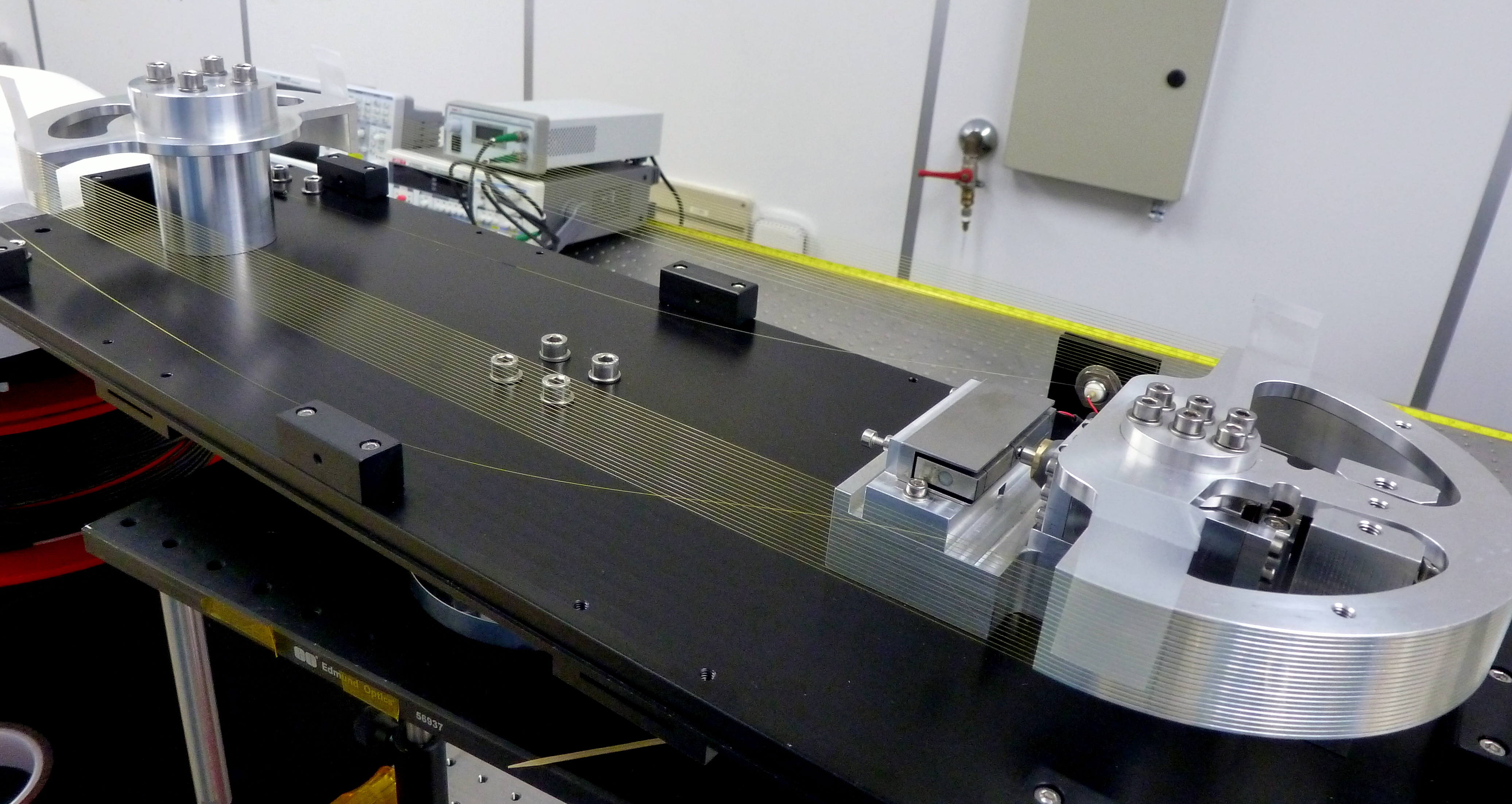}
    \caption{Picture of the final fiber stretcher of NIRPS during integration. Fibers are eventually secured on the grooves with glue patches.}
    \label{fig:stretcher_pic}
\end{figure}

\begin{figure}
    \centering
    \includegraphics[width=1.0\textwidth, trim={0 0.5cm 0 0},clip]{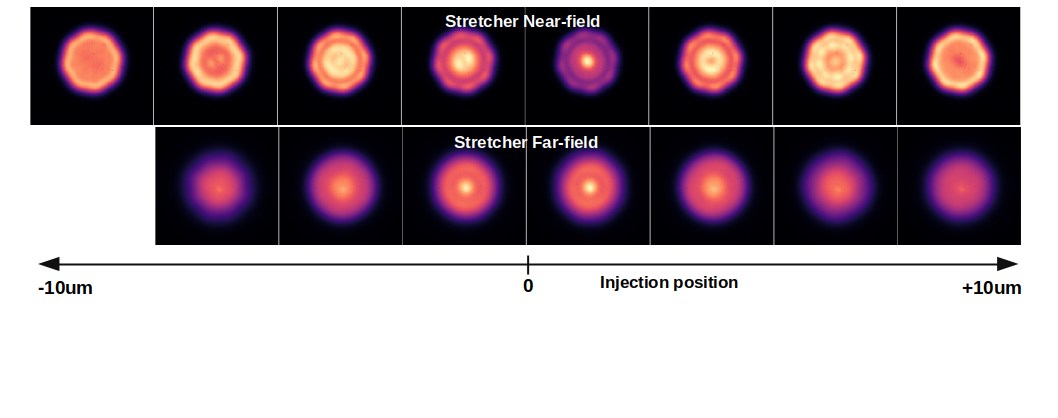}
    \caption{A few images of the stretcher fiber illumination, to be compared to Fig.~\ref{fig:oct29_field}. From left to right the fiber entrance is scanned with a point source, with 1310nm filter.}
    \label{fig:stretcher_field}
\end{figure}

The fiber stretcher (Fig.~\ref{fig:stretcher_pic}) is a device whose primary goal is to decrease the impact of modal noise in the few mode fibers, by quickly modulating phase between modes and homogenizing the output illumination. This allows to significantly reduce the impact of external factors (like varying fiber stress or temperature gradients) and stabilize the cross-correlation function between the measured spectrum and the mask from one observation to another, so as to avoid drifts/noise by more than 1m/s.
\begin{itemize}
    \item \textbf{Phase modulation --} The working principle of the stretcher is to dynamically modulate the phase between modes, which present inter-modal dispersion of $\sim 10^{-4} - 10^{-3}$ (or a beat length in the 'few mm range'). Doing so, it averages the speckle pattern and generate over short periods of time a figure equivalent to the incoherent sum of the excited modes. A phase modulation of 2$\pi$ between any set of two modes requires a few mm stretching. However, the beat length between different pairs of modes being different, the higher the phase modulation amplitude, the better the result.
    \item \textbf{Mode scrambling --} Because of the fiber routing around two half cylinders, modes are slightly disturbed and can transfer energy to others. Hence the stretcher presents scrambler properties by distributing energy over more modes than at its entrance. Such effect was a lot clearer on the final stretcher with octagonal fibers, than on the very first prototype on round fibers. The rotational asymmetry of octagonal fibers leads to finer effective index difference between modes, hence facilitating transfer of energy between modes, as well as making them more sensitive to external disturbance (the cylinder routing here).
\end{itemize}
Fig.~\ref{fig:stretcher_field} shows a few images of the stretcher NF and FF. The benefit of the two previously described effects is clearly visible when compared to Fig~\ref{fig:oct29_field}, although it does not allow to uniformly fill all the modes of the fiber.

The final stretcher (Fig.~\ref{fig:stretcher_pic}) is made of 20 spools, assembled on half-cylinders (R=75mm) on which grooves have been drawn to keep the fiber in place. The stretch is performed with a piezo-amplified actuator, allowing up to 250\:$\mu m$ of stretch per section. We end up with a total of 7mm stretch for the whole assembly, due to a few loose sections. This stretch is equivalent to a temperature change of $\ge$\:50K over 25m fiber. It is modulated at a frequency of 1\:Hz with a sine function to limit shocks and avoid premature use or breakage of the fiber\footnote{Note that the first prototype was operated continuously for 1\:year at a modulation frequency of 30Hz without showing signs of fatigue. It was stopped when we started to work on the 2nd prototype.}. Also, the stretcher is integrated within a 60m long fiber ($\sim$ 20m devoted to the stretcher routing), without connectors, to not introduce fiber-to-fiber misalignment (lateral or rotational) and break the upstream scrambling. The transmission of the stretcher (fiber transmission excluded) is $\ge 90\%$. No additional transmission loss was observed when piezo was stretching.

\subsection{Double scrambler}
\label{sec:double_scrambler}
Optical fibers (and even more non-circular ones) are known for their excellent scrambling properties in the near-field, which led to the very successful concept of double scrambler to exchange near- and far-field properties, and scramble both efficiently. We have seen in Fig.~\ref{fig:oct29_field} that few-mode fibers are not behaving the same way: changing injection conditions very clearly changes the output near-field illumination. Also, the near-field and far-field of few-mode fibers are quite similar in energy distribution and are much less independent from each other.
This observations questioned the benefit of using a double scrambler in NIRPS. In addition, because of mismatch of modes between the near- and far-field, it is expected\textit {1)} to get decreased coupling efficiency compared to a standard feedthroug: (simulations suggest losses $\sim$20\% in the case of a double scrambler); \textit{2)} to actually create some modal noise.
To test this, we designed a double scrambler and a standard feedthrough for NIRPS, and characterized them both at different wavelengths. Note that the input of those device was a single fiber, and not the stretcher prototype presented in previous section, which was still under preparation.

\begin{figure}
    \centering
    \includegraphics[width=0.6\textwidth]{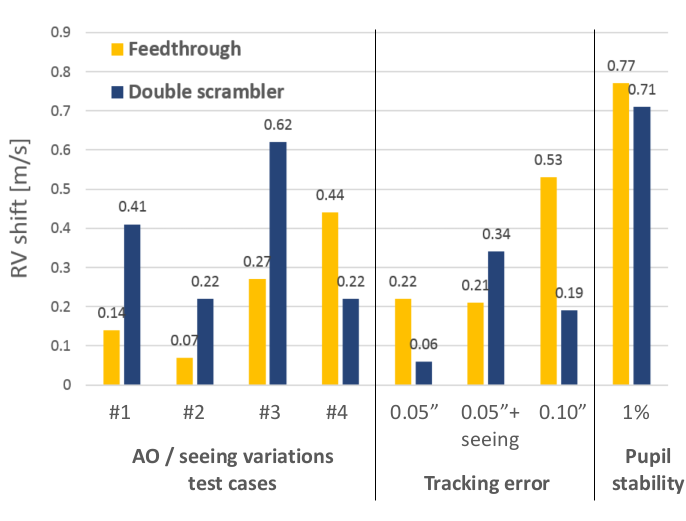}
    \caption{Double scrambler performance compared to a standard feedthrough.}
    \label{fig:double_scrambler}
\end{figure}

One of the benefit of the double scrambler being the gain in grating illumination stability, a definitive comparison was to inject the NF and FF measurements in the spectrograph design, as presented in Sect.~\ref{sec:spectro_sim}. Fig.~\ref{fig:double_scrambler} compares the performance of both device for different test cases, enumerated from left to right:
\begin{enumerate}
    \item AO scrambling with '1-r' law and scanning radius of 0.2";
    \item AO scrambling with 'r' law and scanning radius of 0.2";
    \item AO scrambling with 'uniform' law and scanning radius of 0.2";
    \item AO scrambling with 'uniform' law and scanning radius of 0.4". Scanning is performed outside the fiber core, and might be consider to first order as a 'seeing limited' case.;
    \item Same as 3), with guiding error of 0.05";
    \item Same as 4), with guiding error of 0.05";
    \item Same as 3), with guiding error of 0.1";
    \item Same as 3), with pupil illumination change of 10\% in diameter (this change was forced numerically and not the result of a different scrambling for instance);
\end{enumerate}
There is no clear winner here. A standard feedthrough seems to perform better under smaller modulation radii (tests 1 to 4), but on the other hand appears more sensitive to guiding errors (tests 5 to 7). Both device react equally to change in fiber illumination.

The double scrambler was ultimately implemented in NIRPS because at the time of measurements and simulations, the design was nearly finalized.

\subsection{Summary for HAF / OCT29}
\begin{figure}
    \centering
    \includegraphics[width=0.7\textwidth]{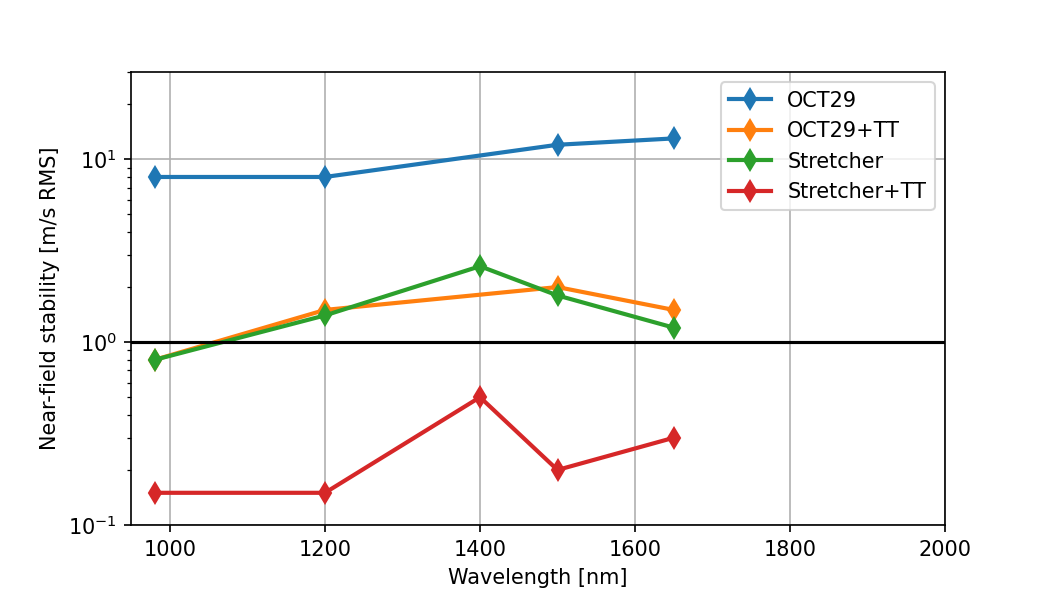}
    \caption{Summary of near-field stability measurement at different wavelengths and for different set-up.}
    \label{fig:scrambling_summary}
\end{figure}

Fig.~\ref{fig:scrambling_summary} summarizes the performance of the OCT29 fiber presented in this section. From the pure RV stability point of view, it appears that the stretcher or the AO scrambling provide each a gain of $\sim$10 on the stability, down to about 1m/s. When combined, they accumulate to first order (thanks to their different action on fiebr modes) and stability drops to the few 10 cm/s level. The measurements and spectrograph simulations do not point to a clear benefit of a double scrambler in the few-mode regime.

\section{The HEF / REC33x132 fiber}
\begin{figure}
    \centering
    \includegraphics[width=0.95\textwidth]{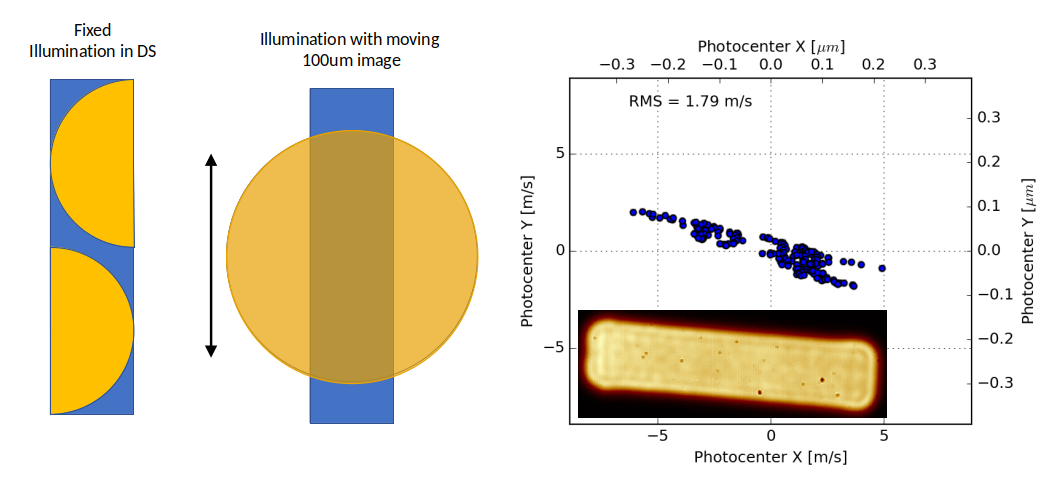}
    \caption{Left: illumination in the double-scrambler with slicer. Middle: moving illumination for the stability test. Right: Photocenter displacement of REC33x132 at $\lambda=1310 \pm 10 nm$, with an image of the near-field when the input source is centered.}
    \label{fig:R33x132_test}
\end{figure}

We could not easily make a mock-up of the slicer part to investigate scrambling performance of the HEF. We instead made a 'worst illumination' scenario test on the rectangular fiber: we used a 500$\mu m$ fiber (magnified by x0.2 by the bench, making a 100 $\mu m$ image), which we moved in front of the rectangular fiber  (Fig.~\ref{fig:R33x132_test}, middle). We scanned with amplitude 40$\times$150 $\mu m$ to evaluate the level of modal noise under a strongly varying illumination condition. The test shows a very symmetric photocenter motion of amplitude $\le$2 m/s along the small side of the fiber (i.e. the main dispersion direction) at $\lambda$=1310\: nm (Fig.~\ref{fig:R33x132_test}, right). 

Considering the following stability of the rectangular fiber illumination in the double-scrambler:
\begin{itemize}
    \item  A scrambling gain $\ge$ 100 at the same wavelength (i.e. illumination photocenter moving by less than 5 $\mu m$, or $\ge$ 150\:m/s);
    \item A thermal stability of the double scrambler opto-mechanics as bad as 10 $\mu m$ in the near-field;
\end{itemize}
would result with photocenter motions at the output of the HEF below 2 m/s × (5$\mu$m + 10$\mu$m)/132$\mu$m = 0.23 m/s. The stability requirement for this fiber was initially set to 5 m/s. Those measurements suggest that such performance should be easily reached.

\section{Conclusion}
A significant effort was made to understand and characterise modal noise in few-mode fibers. Those tests validated the envisioned strategies (use of stretcher and AO scrambling), showing stability better than 0.5m/s. 

A limitation of those tests is nevertheless to have focused solely on the RV stability, averaged over small bandpath (equivalent to one diffraction order of the spectrograph), assuming a very smooth evolution of the modes over 1-10\:nm, given their small number. First commissioning tests on sky have shown the presence of very high spectral resolution features ($\le$0.1A), i.e. about the width of an actual spectral line (see \cite{bouchy_2022a}, these proceedings). Such behavior could only have been observed and studied with a high resolution spectrograph R$\ge$50.000. Such pattern could originate from the double scrambler alignment.

We also underestimated the importance of the modal noise on calibration fibers (with large input illumination), as well as on the HEF, and focused on the AO-fed HA one. The latter, equipped with fiber stretcher and AO scrambling, demonstrates a higher stability of modal noise pattern over a day and for varying telescope pointings.

\bibliography{references} 
\bibliographystyle{spiebib} 

\end{document}